\documentclass[aps,pra,showpacs,reprint]{revtex4-1}
\usepackage{bm}
\usepackage{amsmath}
\usepackage{graphicx}
\usepackage{subfigure}
\usepackage{color}
\usepackage{amsfonts}
\usepackage{amssymb}
\usepackage{amsmath}

\renewcommand{\vec}{\bm}

\def\splitting{\varepsilon}
\def\longit{\parallel}

\begin{document}

\title{Spin splitting of relativistic particles in 3D}

\author{Tihomir G. Tenev}
\affiliation{Department of Physics, Sofia University, 5 James Bourchier Blvd, Sofia 1164, Bulgaria}
\author{Nikolay V. Vitanov}%
\affiliation{Department of Physics, Sofia University, 5 James Bourchier Blvd, Sofia 1164, Bulgaria}%

\date{\today}

\begin{abstract}
The behavior of relativistic particles in an electric and/or magnetic field is considered in the general case when the direction of propagation may differ from the direction of the field.
A special attention is paid to the spin splitting and the ensuing Larmor precession frequency of both neutral and charged particles.
For both neutral and charged particles, the Larmor frequency shows a longitudinal motional red shift.
%
For a neutral particle, there is a dynamical upper bound, which depends on both the mass and the transverse momentum of the particle;
 moreover, the transverse motion leads to a blue shift of the Larmor frequency.
For a charged particle, the longitudinal motional decrease of the spin splitting is determined by the formation of Landau levels and it has no upper limit.
Unlike the nonrelativistic limit, the relativistic spin splitting depends on the Landau levels and decreases for higher Landau levels, thereby signalling the presence of a Landau ladder red shift effect.
\end{abstract}

\pacs{03.67.Ac, 03.65.Pm, 37.10.Ty}

\maketitle

\section{Introduction}
Larmor precession is a well-known effect in nonrelativistic quantum mechanics.
Within the nonrelativistic theory the Larmor precession frequency does not depend on the particle velocity but only on the size of the magnetic dipole moment and the strength of the applied field.
Furthermore, the effect is linear in both the size of the magnetic dipole moment (MDM) and the strength of the field.
There is no upper bound on the Larmor precession frequency: in principle, it increases infinitely as the applied field increases.

Recently~\cite{TenevB}, we have examined the properties of spin splitting of neutral relativistic particles in 1D and showed a deviation of the behavior of the Larmor precession frequency from the nonrelativistic result.
We have shown two notable effects.
First, an upper limit $\splitting=2mc^2$ of the spin splitting in 1D exists, which is independent of the size of MDM or the strength of the applied field.
Second, the spin splitting depends on the particle momentum and velocity, which we have referred to as relativistic motional decrease effect.

Here we explore the properties of spin splitting in 3D for both neutral and charged relativistic particles.
We show that the relativistic motional decrease effect is present, in a modified form, for both neutral and charged particles in 3D.
However, in 3D there is no upper limit for spin splitting for neither neutral nor charged particles.
The reason for this is different in the two cases.

\section{Neutral particle in 3D}
The Hamiltonian of a neutral relativistic particle propagating in a static magnetic and/or electric field reads
\begin{equation}
\hat{H} = c\hat{\vec{\alpha}}\cdot\hat{\vec{p}} + 2\hat{\beta}\hat{S}_{\longit}\left(d E - \mu B\right),
\end{equation}
where $\hat{S}_{\longit}$ is the longitudinal component (along the direction of the applied field) of the spin vector operator in relativistic theory, $E$ is the electric field, $B$ is the magnetic field, $d$ is the size of the electric dipole moment (EDM) and $\mu$ is the size of the magnetic dipole moment (MDM) of the relativistic particle.
The term $2\hat{\beta}\hat{S}_{\longit}\left(d E - \mu B\right)$ lifts the spin degeneracy and the four distinct eigenvalues are
\begin{subequations}\label{EQ:EigenE}
\begin{align}
E_{\pm}^{\uparrow}   &= \pm \sqrt{p_{\longit}^2c^2  + \left(c\sqrt{m^2c^2 + p_{\bot}^2} + \delta\right)^2} ,  \\
E_{\pm}^{\downarrow} &= \pm \sqrt{p_{\longit}^2c^2  + \left(c\sqrt{m^2c^2 + p_{\bot}^2} - \delta\right)^2} ,
\end{align}
\end{subequations}
where $\delta = d E -\mu B$ and $p_{\bot}$ is the transverse component of the particle momentum.
If $p_{\bot}=0$ the eigenvalues~\eqref{EQ:EigenE} reduce to the 1D case considered previously~\cite{TenevB}.

The spin splitting $\splitting = E_+^{\uparrow} - E_+^{\downarrow} = E_-^{\downarrow} - E_-^{\uparrow}$ reads
\begin{align}\label{EQ:SpinSplit}
\splitting &= \sqrt{p_{\longit}^2c^2  + \left(c\sqrt{m^2c^2 + p_{\bot}^2 } + \delta\right)^2} \notag \\
         &- \sqrt{p_{\longit}^2c^2  + \left(c\sqrt{m^2c^2 + p_{\bot}^2 } - \delta\right)^2} .
\end{align}
It depends not only on the longitudinal momentum $p_{\longit}$, along the direction of the applied electric field, but also on the transverse momentum $p_{\bot}$.
However these momentum components enter in the eigenvalues in different manner and lead to qualitatively different behaviors.
Because both the longitudinal momentum $p_{\longit}$ and the transverse momentum $p_{\bot}$ enter quadratically in the spin splitting, the latter does not depend on their signs;
 hence, we shall assume for simplicity that both $p_{\longit}\geqq 0$ and $p_{\bot}\geqq 0$.
Furthermore, because when $\delta$ is replaced by $-\delta$ the spin splitting in Eq.~\eqref{EQ:SpinSplit} changes its sign too, we shall assume without loss of generality that $\delta\geqq 0$;
 then we find from Eq.~\eqref{EQ:SpinSplit} that $\splitting>0$.

Because $\partial_{p_{\longit}}\splitting=-p_{\longit}\splitting/(E_+^{\uparrow}E_+^{\downarrow})<0$
 we conclude that $\splitting(p_{\longit})$ is a monotonically decreasing function of $p_{\longit}$,
 with the maximum value of $\splitting(p_{\longit})$ achieved for $p_{\longit}=0$.
This behavior is the same as in the 1D case~\cite{TenevB}.
The expression for $\splitting_0=\splitting(p_{\longit}=0)$ reads
\begin{equation}
\splitting_0 = \Big|c\sqrt{m^2c^2 + p_{\bot}^2} + \delta\Big| - \Big|c\sqrt{m^2c^2 + p_{\bot}^2} - \delta\Big| .
\end{equation}

\begin{figure}
\begin{center}
\includegraphics[scale=0.8]{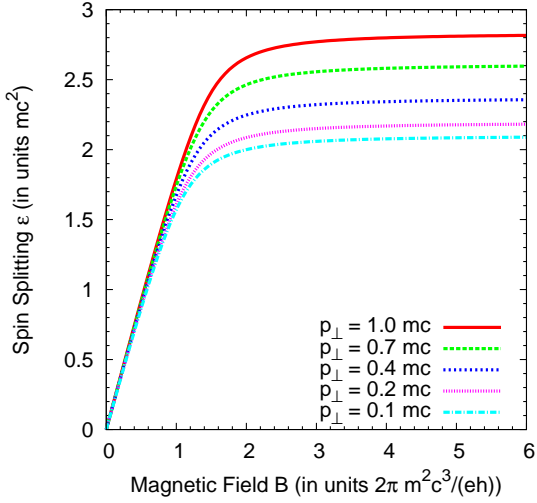}
\end{center}
\caption{(Color online) Spin splitting of a neutral particle vs. interaction energy $\delta$ for different values of the transverse momentum $p_{\bot}$.
The dynamical upper limit $c\sqrt{(mc)^2+p_{\bot}^2}$ increases with the transverse momentum $p_{\bot}$.
The longitudinal momentum is $p_{\longit}=0.3 mc$.}\label{FIG:Fig1}
\end{figure}

There are two distinct regimes depending on the size of the interaction energy $\delta$:
\begin{subequations}
\begin{eqnarray}
\delta < c\sqrt{m^2c^2 + p_{\bot}^2}\  & \  \Rightarrow   \ & \ \splitting_0 = 2\delta ,\label{EQ:A} \\
\delta > c\sqrt{m^2c^2 + p_{\bot}^2}\  & \  \Rightarrow   \ & \ \splitting_0 = c\sqrt{m^2c^2 + p_{\bot}^2} . \label{EQ:B}
\end{eqnarray}
\end{subequations}
Similarly to the 1D model~\cite{TenevB} $\splitting_0$ grows linearly with $\delta$ until a threshold value is reached, after which $\splitting_0$ stays constant.
However, unlike the 1D model~\cite{TenevB}, the threshold value $c\sqrt{m^2c^2 + p_{\bot}^2}$ and the saturation values depend not only on the rest mass energy $mc^2$, but also on the value of the transverse momentum $p_{\bot}$.
Consequently, in the 3D model a global upper limit for the spin splitting does not exist since $p_{\bot}$ may be arbitrarily large.
However, for any fixed $p_{\bot}$, the value $c\sqrt{m^2c^2 + p_{\bot}^2}$ is the maximum value that the spin splitting  $\splitting_0$ can have.
We refer to it as the \emph{dynamical upper limit} for a given transverse momentum $p_{\bot}$.
It is demonstrated in Fig.~\ref{FIG:Fig1}.

\emph{\textbf{Low-speed limit.}} For small values of $p_{\longit}$, the spin splitting \eqref{EQ:SpinSplit} can be expanded in Taylor series versus $p_{\longit}$,
\begin{equation}
\splitting \approx \splitting_0 - \frac{p_{\longit}^2c^2}{2|m^2c^4+p_{\bot}^2c^2-\delta^2|}\splitting_0 \; .\label{EQ:LowSpeedLimit}
\end{equation}
When $p_{\bot}=0$ Eq.~\eqref{EQ:LowSpeedLimit} reduces to the 1D expression \cite{TenevB}.
As $p_{\bot}$ increases the size of the energy shift decreases --- a dependence, which is the opposite to the one on the longitudinal momentum $p_{\longit}$.

\emph{\textbf{High-speed limit.}} For large values of $p_{\longit}$, we find from Eq.~\eqref{EQ:SpinSplit} the high-speed approximation
\begin{equation}\label{EQ:HighSpeedLimit}
\splitting \approx \frac{2\delta\sqrt{m^2c^2+p^2_{\bot}}}{p_{\longit}} .
\end{equation}
For $p_{\bot}=0$, Eq.~\eqref{EQ:HighSpeedLimit} reduces to the high-speed limit in the 1D case~\cite{TenevB}.
Obviously, the spin splitting in this limit increases with the transverse momentum $p_\bot$.

\begin{figure}
\begin{center}
\includegraphics[scale=0.8]{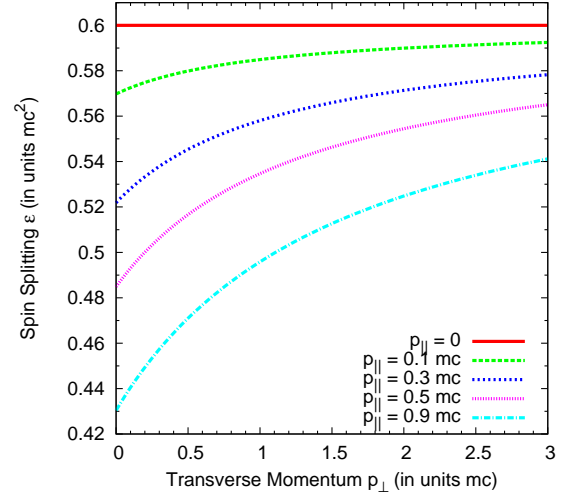}
\end{center}
\caption{(Color online) Spin splitting of a neutral particle vs. the transverse momentum $p_{\bot}$ for different values of the longitudinal momentum $p_{\longit}$
for $\delta =0.3 mc^2$.}\label{FIG:Fig2}
\end{figure}

\begin{figure}
\begin{center}
\includegraphics[scale=0.8]{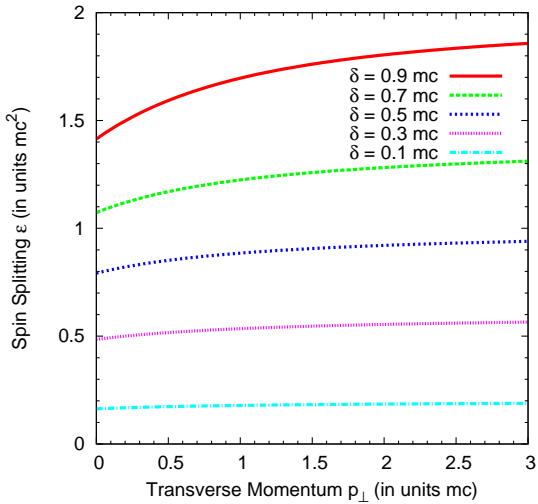}
\end{center}
\caption{(Color online) Spin splitting of a neutral particle vs. the transverse momentum $p_{\bot}$ for different values of the interaction energy $\delta$ for $p_{\longit}=0.5 mc$.}\label{FIG:Fig3}
\end{figure}

Because 
${\partial\splitting}/{\partial p_{\bot}} = {p_{\bot}} \left[\delta_{\longit}\left(E^{\uparrow} + E^{\downarrow}\right) - \splitting \right] / (E^{\uparrow}E^{\downarrow}) > 0$,
the spin splitting $\splitting$ is a monotonically increasing function of $p_{\bot}$.
This is also evident in Eqs.~\eqref{EQ:LowSpeedLimit} and \eqref{EQ:HighSpeedLimit}, and Figs.~\ref{FIG:Fig2} and \ref{FIG:Fig3}.
We refer to this behavior as \emph{transverse motional increase} for the spin-splitting $\splitting$, and \emph{transverse motional blue shift} for the associated Larmor precession frequency $\omega_L=\splitting/\hbar$.

This effect is illustrated in
 Fig.~\ref{FIG:Fig2} for a fixed interaction energy $\delta=0.3mc^2<c\sqrt{m^2c^2+p_{\bot}^2}$.
The line for $p_{\longit}=0$ shows no transverse motional increase, as expected: 
 then the spin splitting is given by Eq.~\eqref{EQ:A}.
If $\delta >c\sqrt{m^2c^2+p_{\bot}^2}$ the spin splitting would not be constant with $p_{\bot}$, but would manifest transverse motional blue-shift effect described by Eq.~\eqref{EQ:B}.
Figure~\ref{FIG:Fig2} shows that the transverse blue shift is more pronounced for larger values of the longitudinal momentum $p_{\longit}$.
Furthermore, the transverse blue-shift effect is larger for larger values of the interaction energy $\delta$, as shown in Fig.~\ref{FIG:Fig3}.

\section{Charged particle in a magnetic field}
The Hamiltonian for a charged particle in an external magnetic field in the Landau gauge reads
\begin{equation}
\hat{H} = \hat{\vec{\alpha}}\cdot(\hat{\vec{p}} - \frac{e}{c}\mathbf{A}) + \beta mc^2 ,\label{EQ:ChargedHam}
\end{equation}
where $\mathbf{A}$ is the vector magnetic potential.
The eigenvalues of the Hamiltonian \eqref{EQ:ChargedHam} are the relativistic Landau levels \cite{Risebor,GreinerRelQM,RelLandauLev,AdvancedSakurai,Zawadzki},
\begin{equation}
E_n^\sigma = c\sqrt{m^2c^2 + p_{\longit}^2 + (2n+1+\sigma)\frac{\hbar e}{c}B} ,
\end{equation}
where $\sigma=\pm1$ denotes ``up'' and ``down'' spin states along the propagation direction.
The spin splitting for the $n$-th Landau level is given by
\begin{align}
\splitting_n &= c\sqrt{m^2c^2 + p_{\longit}^2 + 2(n+1)\frac{\hbar e}{c}B} \notag\\
&- c\sqrt{m^2c^2 + p_{\longit}^2 + 2n\frac{\hbar e}{c}B}.\label{EQ:ChargedSpinSplit}
\end{align}
The spin splitting for a charged particle in 3D is qualitatively different from the expression~\eqref{EQ:SpinSplit} for a neutral particle.
The difference is due to the coupling of the magnetic field to the charge of the particle, which modifies the orbital dynamics by formation of Landau levels.
The relativistic expression \eqref{EQ:ChargedSpinSplit} is also different from the nonrelativistic result,
\begin{equation}\label{EQ:splitting-nonrel}
\splitting_n^{\text{nonrel}} = \frac{\hbar e B}{mc},
\end{equation}
which is readily obtained from the nonrelativistic expression for the Landau levels,
\begin{equation}
E_n^{\text{nonrel}} = mc^2+\frac{\hbar^2k_{\longit}^2}{2m}+\frac{2n+1+\sigma}{2}\frac{\hbar e B}{mc}.
\end{equation}
We note that in the nonrelativistic expression the spin splitting is the same for all Landau levels --- it does not depend on the index $n$.

Because it follows from Eq.~\eqref{EQ:ChargedSpinSplit} that, as for neutral particles, $\partial_{p_{\longit}}\splitting_n=-p_{\longit} \splitting_n/(E_n^{\uparrow}E_n^{\downarrow})<0$,
 we conclude that the spin splitting of a charged particle also exhibits the relativistic motional red-shift effect along the direction of the applied field.
However, this effect shows some quantitative deviations from the motional decrease effect of neutral particles because of the difference in the definitions of $E_n^{\uparrow}$ and $E_n^{\downarrow}$.

\emph{\textbf{Low-speed and low-field limit.}}
The low-speed and low-field approximation of Eq.~\eqref{EQ:ChargedSpinSplit} reads (up to second order in $p_{\longit}$ and $B$)
\begin{equation}\label{EQ:ChargedLowSpeed}
\splitting \approx \frac{\hbar e B}{mc} - \frac{2n+1}{2mc^2}\left( \frac{\hbar e B}{mc} \right)^2 - \frac{1}{2}\frac{\hbar e B}{mc}\left(\frac{p_{\longit}}{mc}\right)^2.
\end{equation}
The first term is the nonrelativistic result \eqref{EQ:splitting-nonrel};
 within it the spin splitting does not depend on the particle longitudinal momentum $p_{\longit}$ and the Landau level index $n$, and the splitting is linear in the magnetic field $B$.
The next terms depend on both $p_{\longit}$ and $n$, and introduce a quadratic dependence on the magnetic field.
In this low-speed limit the relativistic motional decrease effect is embodied in the $p_{\longit}^2$-term.
Obviously, for heavier particles the spin splitting decrease is smaller.

\begin{figure}
\includegraphics[scale=0.8]{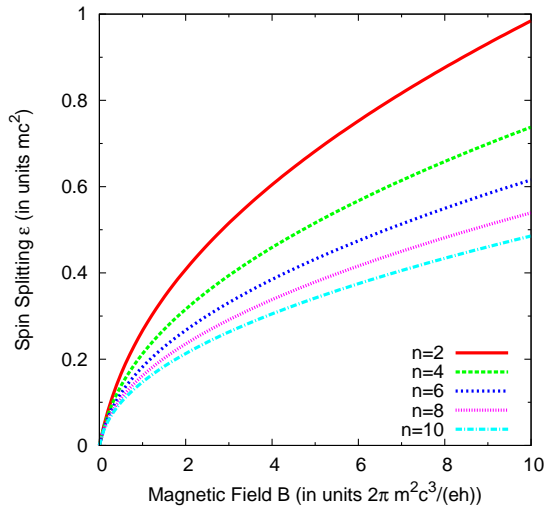}
\caption{(Color online) Spin splitting of a charged particle vs. the magnetic field for different Landau levels $n$ and for $p_{\longit}=0.2mc$.}\label{FIG:Fig4}
\end{figure}

We point out that the $B^2$-term in Eq.~\eqref{EQ:ChargedLowSpeed} also reduces the spin splitting.
This term introduces an $n$-dependence in the relativistic spin splitting: the larger the Landau level index $n$ the greater the red shift.
We refer to this effect as \emph{Landau ladder red shift}.
This effects signals yet another difference from the nonrelativistic result, in which the spin splitting is the same for all Landau levels.
We note that the Landau level index $n$ in some sense substitutes the quantum numbers $p_y$ and $p_z$ of transverse motion (if the $B$-field is along $x$).
We have shown above that for neutral particles the transverse motion leads to a \emph{blue shift} of the spin splitting,
 which is qualitatively the opposite of the effect of $n$ on the spin splitting for charged particles here.
This difference is due to the formation of Landau levels.
The Landau ladder red-shift effect is demonstrated in Fig.~\ref{FIG:Fig4} where it is evident that the spin splitting is smaller for higher Landau levels.

\emph{\textbf{High-speed limit.}}
In the limit of very high speeds ($v\sim c$) the spin splitting \eqref{EQ:ChargedSpinSplit} has the asymptotic behavior
\begin{equation}
\splitting_n(v\sim c) = \frac{\hbar e B}{p_{\longit}}. 
\end{equation}
This expression describes how the spin splitting tends to zero when $v\sim c$.
Notably, it does not depend on the mass of the particle and the Landau levels.
It is proportional only to the magnetic field strength and inversely proportional to the longitudinal momentum.
The behavior vs. $p_{\longit}$ is very similar to the high-speed limit \eqref{EQ:HighSpeedLimit} of the spin splitting of neutral particles.

With respect to the strength of the magnetic field $B$ the spin splitting \eqref{EQ:ChargedSpinSplit} grows monotonically but non-linearly, as it is also seen in Fig.~\ref{FIG:Fig4}.
In contrast to neutral particles, there is no upper bound for the spin splitting versus the magnetic field strength.
However, for a given value of $B$, the nonrelativistic value \eqref{EQ:splitting-nonrel} is always larger than the relativistic one, which is subjected to the motional red shift and the Landau ladder red shift.

\section{Conclusions}
In conclusion, we have investigated the spin splitting of neutral and charged relativistic particles in 3D.
We have shown that for a neutral particle, the upper bound for the 1D case is modified to a \emph{dynamical} upper bound in the 3D case.
The dynamical upper bound depends on both the mass of the particle and the transverse momentum.
We have shown that the transverse motion of the neutral particle leads to a motional increase of the spin splitting and correspondingly, to a transverse motional blue shift for the associated Larmor frequency.
We have shown that the longitudinal motional decrease of the spin splitting is also present for charged particles.
However, there are differences between neutral and charged particles due to the formation of Landau levels in the latter.
Furthermore, we have shown that for charged particles, there is no analog of the transverse motional blue shift.
Instead, the spin splitting depends on the Landau levels and decreases for higher Landau levels.
We have referred to this effect as Landau ladder red shift.

\acknowledgments
This work has been supported by the Bulgarian NSF grant D002-90/08.

%

\end{document}